\documentclass[pra,twocolumn,superscriptaddress,showpacs,10pt,floatfix]{revtex4-1}

\usepackage{blindtext}

\usepackage{amsmath}
\usepackage{amssymb}
\usepackage{amsfonts}
\usepackage{graphicx}
\usepackage{microtype}
\usepackage{accents}

\usepackage[svgnames]{xcolor}
\usepackage[pdftex]{hyperref}
\hypersetup{colorlinks=true,linkcolor=blue,citecolor=ForestGreen,urlcolor=DarkOrchid}

\graphicspath{{/}}

\newcommand{\kff}{$\text{K}_{4,4}$ }

\def\DWnsp{\mbox{DW2000Q}}
\def\ICMnsp{\mbox{PT+ICM}}
\def\ICMLnsp{\mbox{PT+ICM+L}}
\def\HFSnsp{\mbox{HFS}}
\def\DW{\mbox{\DWnsp$\ $}}
\def\ICM{\mbox{\ICMnsp$\ $}}
\def\ICML{\mbox{\ICMLnsp$\ $}}
\def\HFS{\mbox{\HFSnsp$\ $}}
\def\sz{\sigma^{z}}
\def\sx{\sigma^{x}}
\def\H{\mathcal{H}}
\def\G{\mathcal{G}}

\def\hen{H}
\def\king{K}

\newcommand{\nasa}{\affiliation{Quantum Artificial Intelligence Lab., NASA Ames Research Center, Moffett Field, CA 94035, USA}}
\newcommand{\sgt}{\affiliation{Stinger Ghaffarian Technologies Inc., 7701 Greenbelt Rd., Suite 400, Greenbelt, MD 20770}} 
\newcommand{\tamu}{\affiliation{Department of Physics and Astronomy, Texas A\&M University, College Station, Texas 77843-4242, USA}}

\newcommand{\sfi}{\affiliation{Santa Fe Institute, 1399 Hyde Park Road, Santa Fe, New Mexico 87501 USA}}
\newcommand{\oqbit}{\affiliation{1QB Information Technologies (1QBit), Vancouver, British Columbia, Canada V6B 4W4}}

\begin{document}

\title{A deceptive step towards quantum speedup detection}

\author{Salvatore Mandr{\`a}}
\email{salvatore.mandra@nasa.gov}
\nasa
\sgt

\author{Helmut G.~Katzgraber}
\email{hgk@tamu.edu}
\tamu
\oqbit
\sfi

\date{\today}

\begin{abstract}

There have been multiple attempts to design synthetic benchmark problems
with the goal of detecting quantum speedup in current quantum annealing
machines. To date, classical heuristics have consistently outperformed
quantum-annealing based approaches. Here we introduce a class of
problems based on frustrated cluster loops --- deceptive cluster loops
--- for which all currently known state-of-the-art classical heuristics
are outperformed by the \DW quantum annealing machine. While there is a
sizable constant speedup over all known classical heuristics, a
noticeable improvement in the scaling remains elusive. These results
represent the first steps towards a detection of potential quantum
speedup, albeit without a scaling improvement and for synthetic
benchmark problems.

\end{abstract}


\maketitle

\section{Introduction}
\label{sec:intro}

Quantum annealing (QA)
\cite{kadowaki:98,farhi:01,finnila:94,martonak:02,santoro:02,das:08} has
been proposed as a potentially efficient heuristic to optimize hard
constraint satisfaction problems. In principle, the approach can
overcome tall energy barriers commonly found in this class of
optimization problems by exploiting quantum effects, thereby
potentially outperforming commonly-used heuristics that use thermal
kicks to overcome the barriers.  However, despite a significant
effort by the scientific community towards an optimization technique
that, in principle, relies on quantum effects, it is still unclear
whether quantum speedup is actually achievable using analog
transverse-field quantum annealing approaches.

There have been multiple attempts to define quantum speedup
\cite{ronnow:14a,mandra:16b}, as well as quantify any ``quantumness''
and problem-solving efficacy of current commercially-available quantum
annealers
\cite{johnson:11,dickson:13,boixo:14,katzgraber:14,ronnow:14a,katzgraber:15,
heim:15,hen:15a,albash:15a,martin-mayor:15a,king:15,marshall:16,denchev:16,king:17,albash:17}.
However, to date, any convincing detection of an improved scaling of
quantum annealing with a transverse field over state-of-the-art
classical optimization algorithms remains elusive.  The increase in
performance of quantum annealing machines in the last few years has
resulted in an ``arms race'' with classical optimization algorithms
implemented on CMOS hardware. The goal post to detect quantum speedup
continuously keeps moving and has resulted in a renaissance in classical
algorithm design to optimize hard constraint-satisfaction problems.

A key ingredient in the detection of quantum speedup is the selection of
the optimization problems to be used as benchmark. Ideally, one would
want a real-world industrial application where the time to solution of
the quantum device scales better than any known algorithm with the size
of the input.  However, such application problems are not
suitable for present-day quantum annealers, either because they
require more variables than currently available or because precision
requirements cannot be met by current technologies. Random spin-glass
problems have been shown to be too easy to detect any scaling
improvements \cite{katzgraber:14,katzgraber:15}.  As such, efforts have
shifted to carefully-designed synthetic problems.  While some studies
focus on post-selection techniques \cite{katzgraber:15}, others
focus on the use of planted solutions \cite{hen:15a,king:15,king:17}, or
the use of gadgets \cite{albash:17}. Unfortunately, however, in
all planted problems \cite{hen:15a,king:15,king:17} used to date, as well as
problems that use gadgets \cite{denchev:16}, the underlying logical
structure is easily decoded and the underlying problem trivially solved,
sometimes even with exact polynomial methods \cite{mandra:16b}.
Therefore, in the quest for quantum speedup, an important step is to
design problems where no variable reduction or algorithmic trick can be
exploited to reduce the complexity of the problem. Ideally, the
benchmark problem should be hard for a small number of variables and
``break'' all known optimization heuristics.

In this work we introduce a class of benchmark problems designed for \DW
quantum annealers whose logical structure is not directly recognizable
and whose typical computational complexity can be tuned via a control
parameter that tunes the relative strength of inter- vs intracell
couplers in the \DW Chimera \cite{bunyk:14} topology.  Note that this
approach can be easily generalized to other topologies.  We demonstrate
that for a particular setting of the control parameter where the ground
state of the virtual problem cannot be decoded, the D-Wave Systems
Inc.~\DW quantum annealer outperforms all known classical optimization
algorithms by approximately two to three orders of magnitude. More
precisely, we compare against the two best heuristics to solve
Ising-like problems on the \DW Chimera topology, the Hamze-de
Freitas-Selby (\HFSnsp) \cite{hamze:04,selby:14} and parallel tempering
Monte Carlo with isoenergetic cluster moves (\ICMnsp) \cite{zhu:15b}
heuristics.  Although we were not able to identify the optimal annealing
time given the hard limit of $1\,\mu s$ as minimum annealing time in the
\DW device, the scaling is comparable and the speedup persists for
increasing system sizes.  Therefore, we present the first steps towards
the detection of potential quantum speedup, however, for now, without a
noticeable scaling improvement. Problems with tunable complexity as the
ones shown here, combined with a careful statistical analysis,
bulletproof definitions of quantum speedup, the inclusion of power
consumption in the analysis, as well as the use of the currently
best-available heuristics are key in the assessment of the performance
of quantum-enhanced optimization techniques.

The paper is structured as follows. In Sec.~\ref{sec:tech} we present
some technical details of this study. In particular, we introduce the
benchmark problems used in Sec.~\ref{sec:model}. Results are outlined
in Sec.~\ref{sec:results}, followed by a discussion and concluding
remarks.

\section{Technical Details}
\label{sec:tech}

The \DW quantum annealer is designed to optimize classical problem
Hamiltonians of the quadratic form
\begin{equation}
\H_{\rm P} = \sum_{\left(i,\,j\right)\in\G}J_{ij}\sz_i\sz_j + \sum_i h_i \sz_i , 
\end{equation}
where $\G$ is known as Chimera graph \cite{bunyk:14} constructed of a
two-dimensional lattice of fully-connected \kff cells.  The couplers
$\{J_{ij}\}$ and biases $\{h_i\}$ are programmable parameters that
define the optimization problem to be studied. Although the \DW Chimera
architecture graph has been kept fixed since the first commercial
generation of the machine, the number of qubits doubled almost every two
years. At the moment, the latest \DW chip counts $2023$ working
flux-qubits and $5871$ working couplers. To minimize the cost function
$H_{\rm P}$, the \DW quantum chip anneals quantum fluctuations driven by
a transverse field of the form
\begin{equation}
\H_{\rm D} = \sum_i \sx_i .
\end{equation}
More precisely, the annealing protocol starts with the system
initialized to a quantum paramagnetic state.  Then, the amplitude of
$H_{\rm D}$ is slowly reduced while the amplitude of the problem
Hamiltonian $H_{\rm P}$ is gradually increased. If the annealing is slow
enough, the adiabatic theorem \cite{morita:08} ensures that the quantum
system remains in its instantaneous lowest energy state for the entire
annealing protocol. Therefore, (close-to) optimal configurations for
$H_{\rm P}$ can be retrieved by measuring the state of the qubits
along the $z$-basis at the end of the anneal.

Given its intrinsic analog nature, combined with the heuristic properties of
quantum annealing, the \DW device is only able to find the optimum of a cost
function up to a probability $p$. Indeed, fast annealing in proximity of level
crossings \cite{kadowaki:98, farhi:01, santoro:06}, as well as quantum
dephasing effects \cite{amin:09a, dickson:13, albash:15b}, thermal excitations
\cite{wang:16z,nishimura:16,marshall:17z} and programming errors
\cite{mandra:15, katzgraber:15}, can lead to higher energy states of $H_{\rm
P}$ at the end of the anneal.  A commonly-accepted metric is the
time-to-solution (TTS). The TTS is defined as the time needed for a heuristic,
either classical or quantum, to find the lowest energy state with $99\%$
success probability, that is:
\begin{equation}\label{eq:tts}
\text{TTS} 
	= t_\text{run} R_{99} 
	= t_\text{run}\cdot\frac{\log(0.01)}{\log(1-p)},
\end{equation}
where $t_\text{run}$ is either the running time (for a classical
heuristic) or the annealing time (for the \DW quantum chip) and
$R_{99}$ is the number of repetitions needed to reach the desired
success probability \cite{ronnow:14a}. In this work we analyze 
the TTS as a function of the number of input variables in the problem.

\section{Synthetic Benchmark Problems}
\label{sec:model}

In this Section we outline and discuss a new synthetic benchmark we call
``deceptive cluster loop'' (DCL) problems based on traditional
frustrated cluster loop problems. However, DCL problems have a tunable
parameter that for particular values hides the underlying logical
structure of the planted problem, thus ``deceptive.''

\subsection{Traditional frustrated cluster loop problems}

Based on the fact that it is typically hard for agnostic optimization
algorithms to find the lowest energy state of very long frustrated
chains, the frustrated cluster loop {\`a} la Hen (\hen-FCL) is a
random model that has been proven to be hard for many classical
heuristics \cite{hen:15a}. The idea is simple: Given $\mathcal{G}$, an
arbitrary connectivity graph for the problem Hamiltonian $\H_{\rm P}$,
and two parameters $\alpha$ and $R$, \hen-FCL instances are
constructed as follows: 
\begin{enumerate}

\item Generate $M = \alpha n$ loops on the graph, where $n$ is the
number of nodes in $\mathcal{G}$. Loops are constructed by placing
random walkers on random nodes (tails are eliminated once random walkers
cross their own path).

\item For each loop $k$, assign $J^k_{ij} = -1$ to all the corresponding
couplings but one randomly chosen one, for which the value $J^k_{ij} =
+1$ is assigned instead.

\item The final Hamiltonian is then constructed by adding up all the loop
couplings, i.e.,
\begin{equation}
H_P = \sum_{(i,\,j) \in \mathcal{G}}\sum_{k = 1}^M
J^k_{ij}\,\sz_i\sz_j .
\end{equation}
The instance is discarded if there is a coupling such that
\begin{equation}
\left|\sum_{k=1}^M J^k_{ ij}\right| > R .
\end{equation}

\end{enumerate}
The parameters $\alpha$ and $R$ correspond to the density of
``constraints'' and to the ``ruggedness'' of the \hen-FCL problem,
respectively.

Although the \hen-FCL problems can be, in principle, directly
generated for the Chimera graph \cite{hen:15a}, in a recent paper
\cite{king:17}, King {\em et al.}~have chosen a different approach
(called here \king-FCL) that can be divided into two steps:
\begin{enumerate}

\item All couplings inside a \kff unit cell of the Chimera structure are
set to be ferromagnetic, i.e., $J_{ij} = -1$, $\forall i,j \in \text{K}_{4,4}$.
Because the unit cells are fully-connected, all the physical qubits
within a single cell are forced to behave as a single virtual qubit.
This process generates a two-dimensional $L \times L$ lattice with open
boundary conditions of these virtual variables. Here, $L^2$ is the
number of \kff cells on the Chimera graph with $N = 8L^2$ physical
variables (qubits) and $L^2$ virtual variables.

\item The embedded instances are then generated on the virtual lattice
with a given $\alpha$ and $R$.
\end{enumerate}

These \king-FCL problems \cite{chicken} have considerably fewer
(virtual) variables than other benchmarks, but have proven to be
computationally difficult for many heuristics, in particular the \HFS
and \ICM solvers \cite{king:17}. We emphasize, however, that the virtual
problem is planar and can therefore be solved in polynomial time using
minimum-weight-perfect-matching techniques
\cite{kolmogorov:09,mandra:17a}. As such, any speedup claims based on
these problems have to be taken with a grain of salt.

\subsection{Deceptive cluster loop benchmark problems}

Inspired by the \king-FCL benchmark problems, we have developed a new
class of problems we call deceptive cluster loops. Although the ground
state of the problem cannot be planted and therefore has to be computed
with other efficient heuristics, we show that while the \DW device
maintains its performance for this class of problems, all other known
heuristics struggle with solving these instances. In addition, the
virtual problem cannot be easily decoded, i.e., the problems cannot be
solved in polynomial time or with other clever approximations that
exploit the logical structure \cite{mandra:16b}.

The structure of the DCL problems can be summarized as follows: Starting
from an embedded \king-FCL instance, all the inter-cell couplers between 
\kff cells are multiplied by a factor $\lambda$, whereas all intra-cell
couplers have magnitude $1$. To make the problem easier and allow for a
better analysis of the computational scaling, we break the loops
removing one of the ferromagnetic couplings.

One of the main feature of the proposed DCL problems that
distinguishes them from other FCL-like models \cite{king:15,king:17,albash:17}
is the presence of two distinct limits for small and large $\lambda$.
For small $\lambda$, i.e., in the limit of weak inter-cell couplings,
each unit cell results to be strongly connected and therefore, behaves
like a single virtual variable. In particular, when $\lambda \to 1$,
the DCL problems are equivalent to \king-FCL problems. The
corresponding Ising model has a two-dimensional \emph{planar} square
lattice as the underlying graph and therefore can be solved in
polynomial time \cite{mandra:17a}. On the other hand, in the limit of
large $\lambda$, i.e., either horizontal or vertical chains that go
across different unit cells become strongly coupled. Recalling that
the Chimera graph is bipartite, there must exist a gauge
transformation, i.e., a combination of spin flips in one of the two
partitions, such that all the inter-cell couplings can be fixed to be
ferromagnetic. Since in the limit of $\lambda \gg 1$ the energy
contribution of intra-cell couplings is small compared to the energy
contribution of inter-cell couplings, the ground state likely has all
the chains polarized and therefore, they can be treated as single
``virtual'' variables and the corresponding virtual model is the
fully-connected bipartite model \cite{venturelli:15a}. For
intermediate values of $\lambda$, the DCL problems become a nontrivial
combination of the two limits and therefore, optimal states cannot be
mapped onto either virtual models. The effect becomes most pronounced
when $|J_{ij}|$ for the inter-cell couplers is comparable to the
connectivity of the intra-cell variables, i.e., $|J_{ij}| \approx 5$
-- $6$ for the current D-Wave Chimera architecture, where the local
intra-cell environment felt by a variable in the \kff cell competes
with the strength of the inter-cell couplers.

From a physical point of view, the DCL problems have another important
property which makes them interesting in their own right: By
continuously changing the scaling parameter $\lambda$, it is possible
to modify the critical spin-glass temperature from $T_c = 0$ ($\lambda
\sim 1$) \cite{katzgraber:14}, to $T_c > 0$ ($\lambda \gg 1$)
\cite{venturelli:15a}. Therefore, it would be interesting to
understand the nature of the spin-glass phase for intermediate
$\lambda$ where the system is neither planar or fully-connected
\cite{mandra:18}.

\begin{figure*}
  \includegraphics[width=\columnwidth]{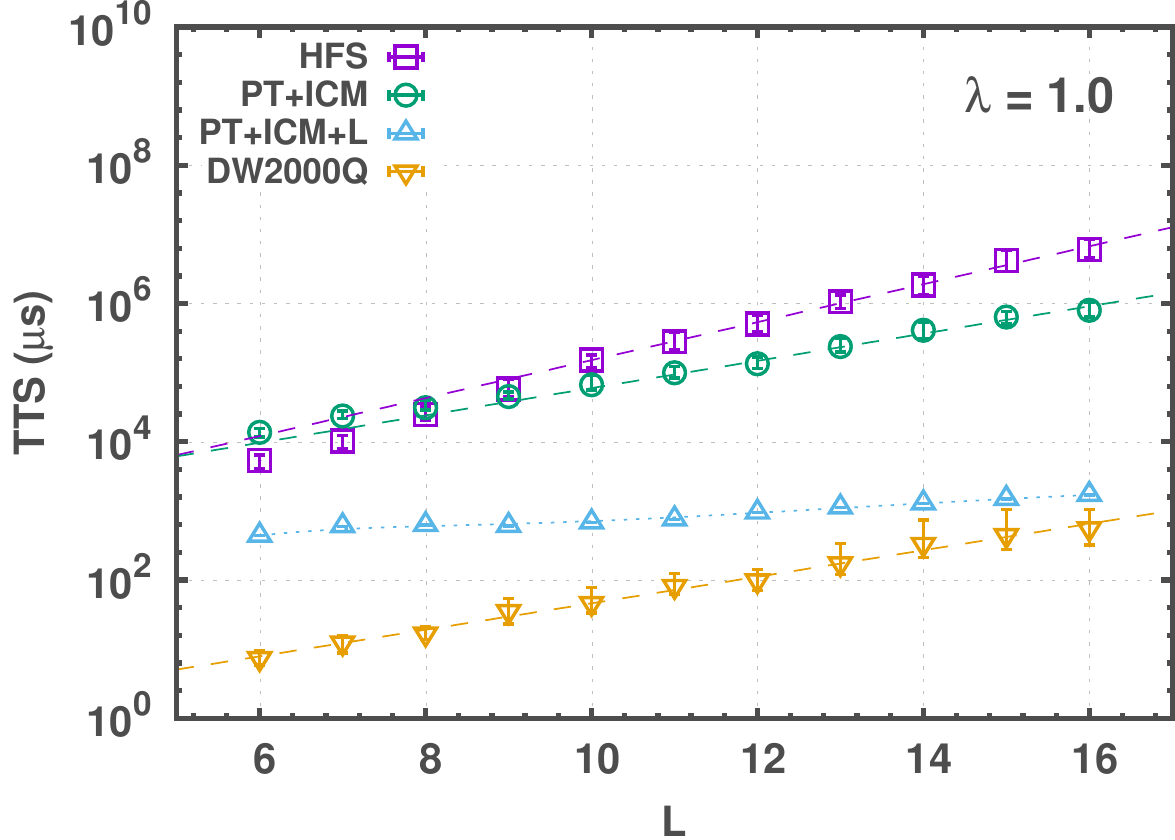}\hspace{1em}
  \includegraphics[width=\columnwidth]{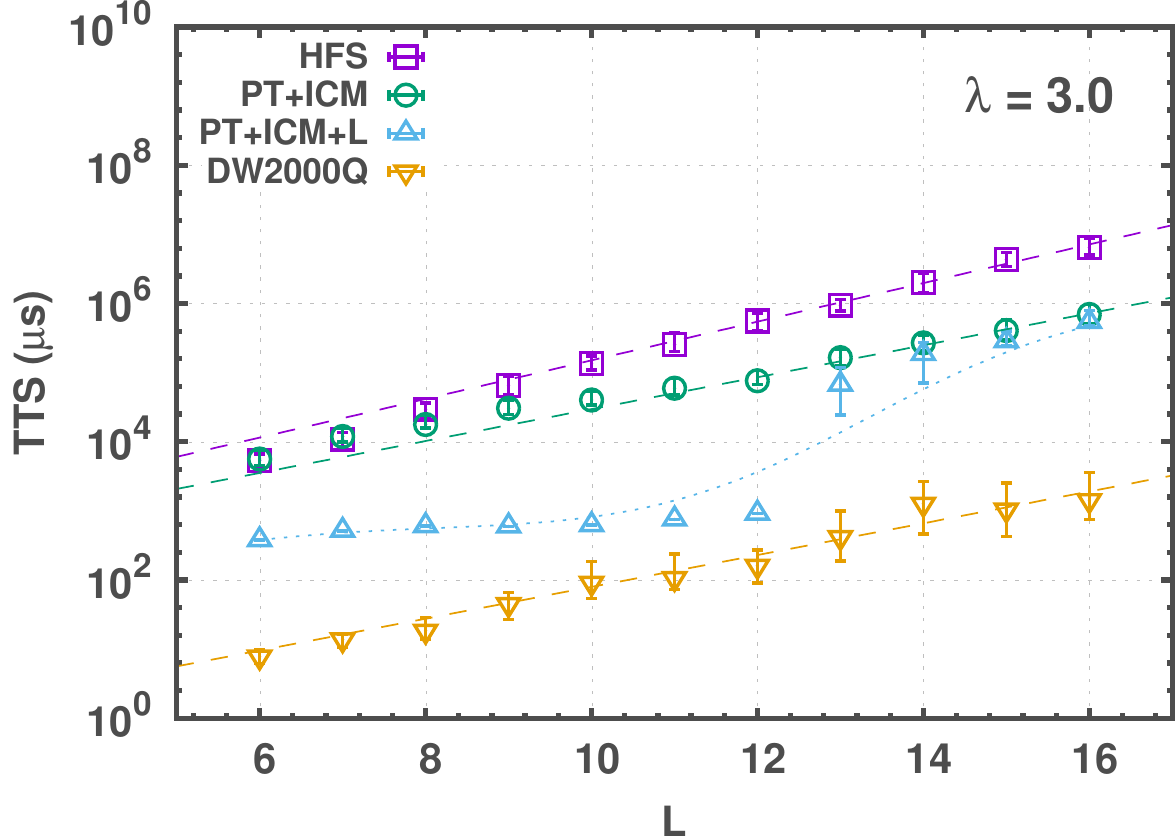}\vspace{1em}\\
  \includegraphics[width=\columnwidth]{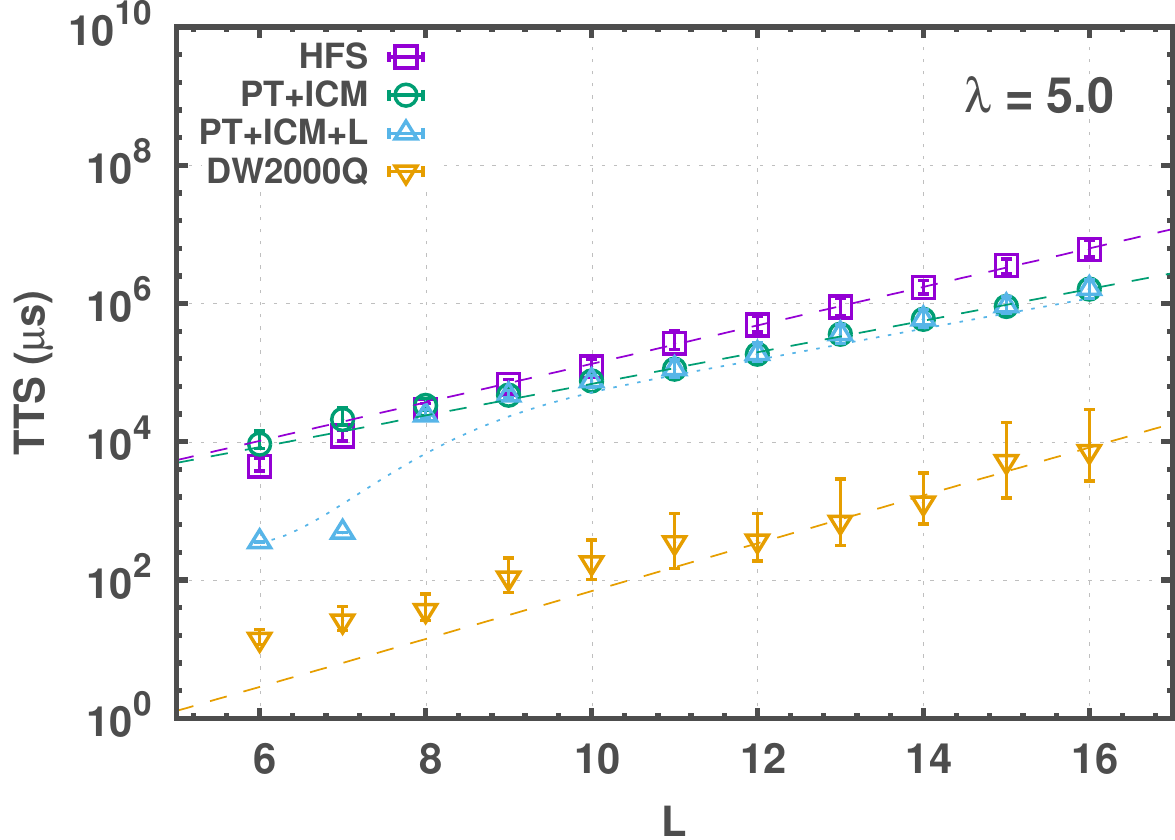}\hspace{1em}
  \includegraphics[width=\columnwidth]{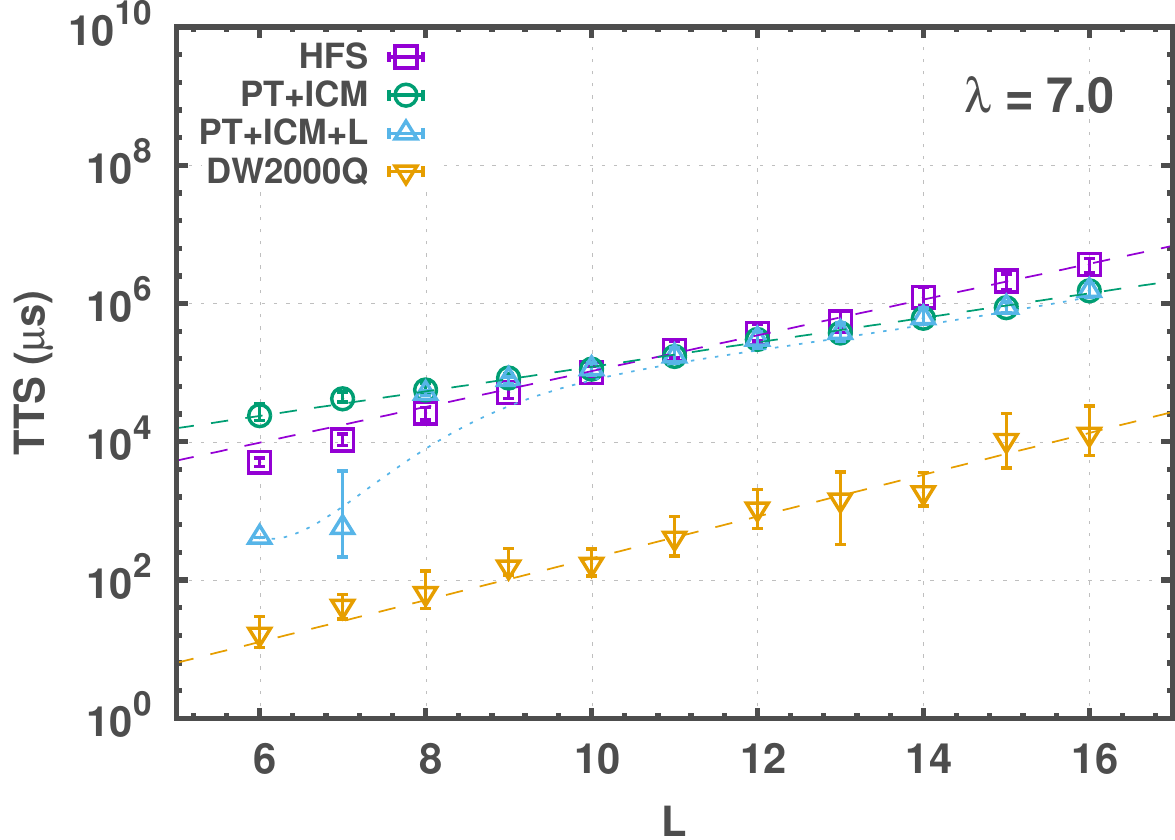}
  \caption{\label{fig:tts_comp_scaling} Time-to-solution (TTS) for the
  parallel tempering isoenergetic cluster method (\ICMnsp), the Hamze-de
  Freitas-Selby (\HFSnsp) heuristic, as well for the D-Wave 2000Q
  (\DWnsp) quantum chip. All data points are for fixed scaling $\lambda$
  while changing the linear problem size $L$. For this analysis, we also
  used a modified \ICM algorithm (\ICMLnsp) to take advantage of the
  knowledge of the virtual ground state for both small ($\lambda \to 1$)
  and large ($\lambda \gg 1$) scaling limits (see main text for more
  details). For \DWnsp, \ICMnsp~and \HFSnsp, fits are obtained by
  considering the last $5$ points only (fit parameters are reported in
  the Appendix). For \ICMLnsp, points are fit with a Bezier curve to
  guide the eye.}
\end{figure*}

\begin{figure}
  \includegraphics[width=\columnwidth]{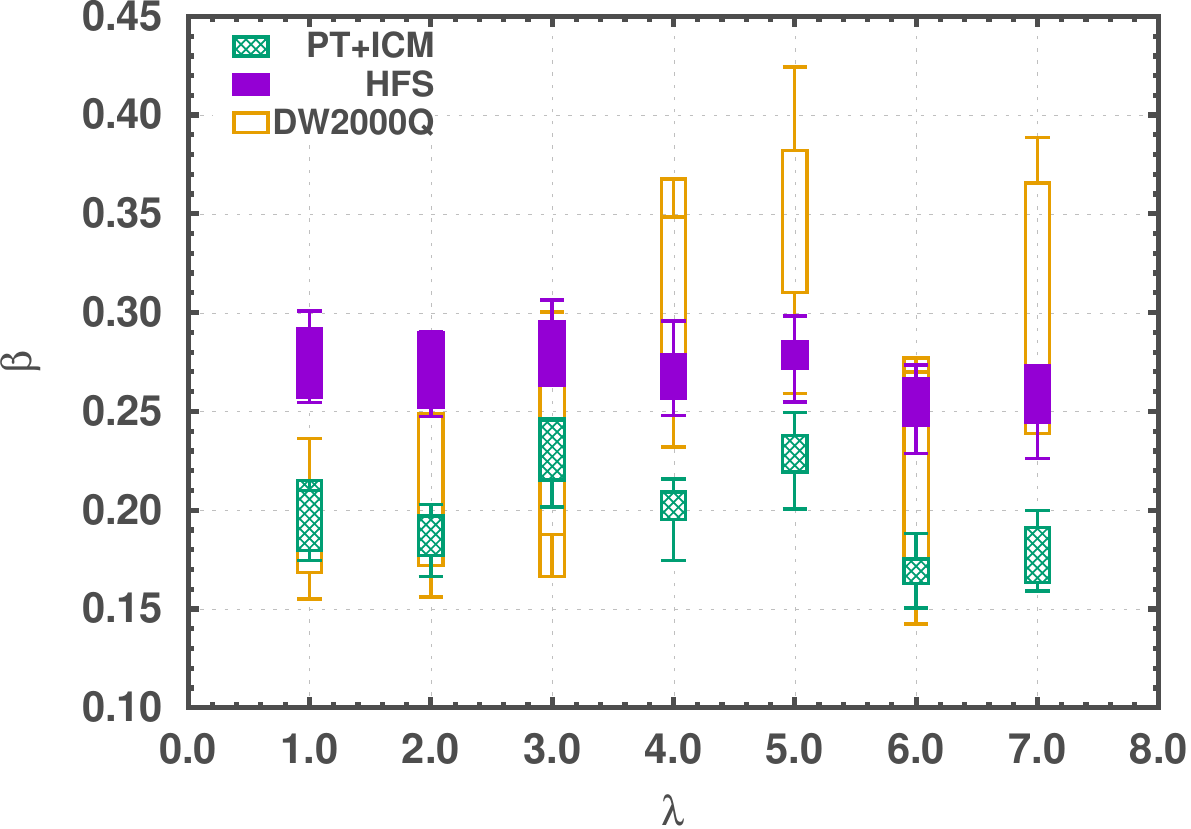}\vspace{1em}\\
  \includegraphics[width=\columnwidth]{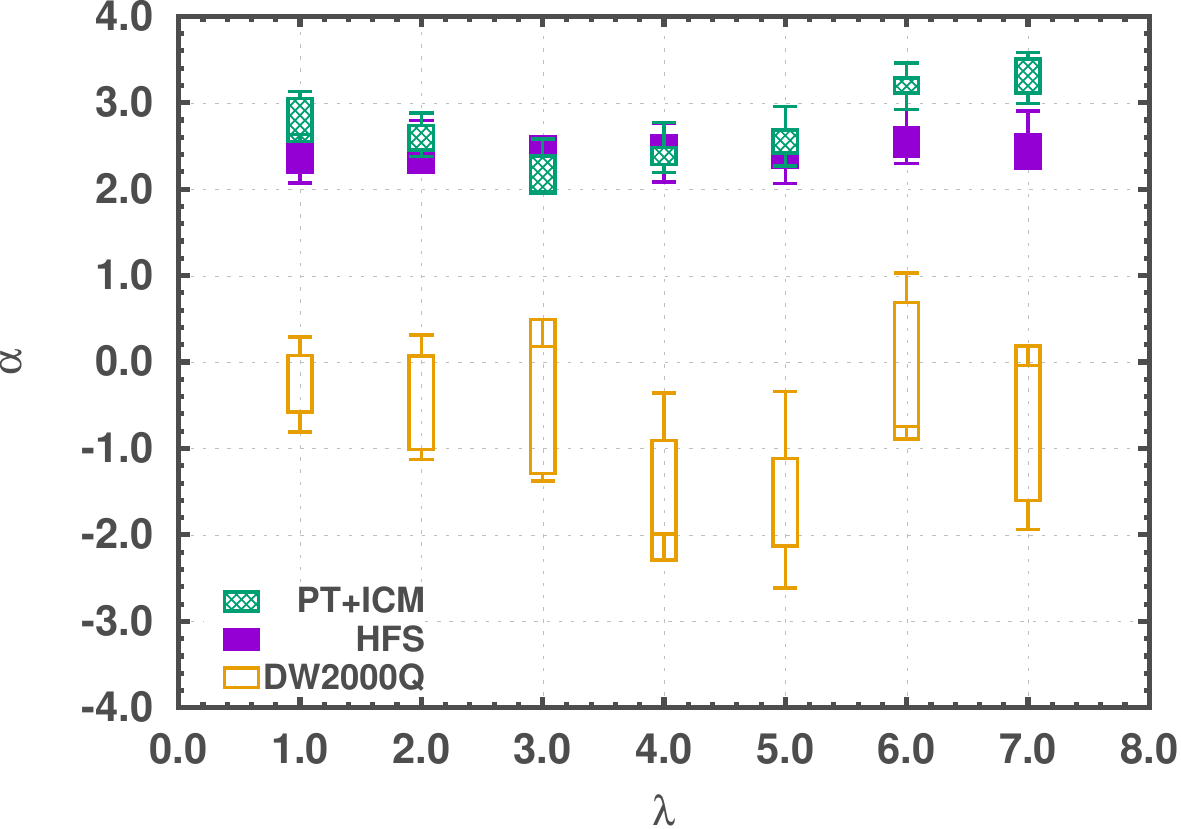}
  \caption{\label{fig:tts_comp_alpha_beta_icm_hfs} Exponential fit
  parameters for the time-to-solution (TTS) of the form
  \mbox{$\log_{10}\text{TTS} = \beta\, t_\text{run} + \alpha$}, for \DW,
  \ICM and \HFSnsp. While the computational scaling parameters $\beta$
  (top) are not significantly different, \DW is over $100$ times faster
  than \ICM and \HFSnsp~when analyzing the prefactor $\alpha$ (bottom) .
  The linear regression is computed by considering only the last five
  linear sizes $L$ [see Fig.~(\ref{fig:tts_slope_scaling_all})].}
\end{figure}

\section{Results}
\label{sec:results}

In this Section, we compare the \DW quantum chip against two of the
fastest classical heuristics for Chimera Hamiltonians, namely the
Hamze-de Freitas-Selby (\HFSnsp) heuristic and the parallel tempering
isoenergetic cluster method (\ICMnsp) \cite{sqa}. Both \HFS and \ICM have been
modified to correctly compute TTS as described in Eq.~(\ref{eq:tts}).
Moreover, \ICM has been further optimized to exploit the knowledge of
the virtual ground states in both limits of small ($\lambda \to 1$) and
large ($\lambda \gg 1$) scaling (referred to as \ICMLnsp). In
particular, $\text{TTS}_\text{\ICMLnsp}$ is computed by running \ICM
from either an initial random state or from one of the two virtual
ground states and then taking the minimum value. For each linear size
$L$, we generated $100$ DCL instances with parameters $\alpha = 0.24$
and $R = 1$ (instances at different $\lambda$ have been obtained by
properly rescaling the inter-cell couplings). In all plots, points
represent the median of the distribution while the error bar correspond
to the $35\%$--$65\%$ percentiles. If not otherwise indicated, \DW
annealing time has been fixed to the minimum allowed, namely $5\,\mu s$.
Simulation parameters for the classical heuristics are listed in the
Appendix.

\begin{figure}
  \includegraphics[width=\columnwidth]{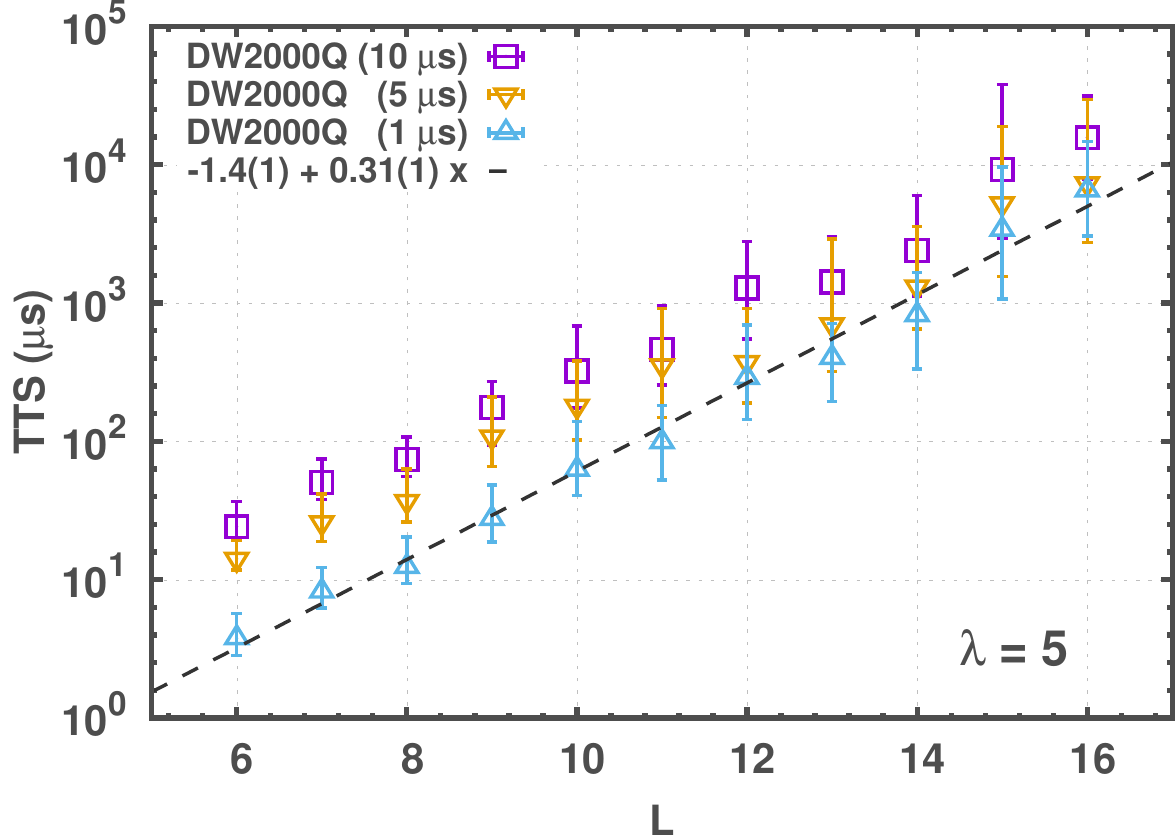}
  \caption{\label{fig:tts_dw_varying_ann}Comparison of the
  time-to-solution (TTS) for \DW by varying annealing time. Despite the
  fact that we are not able to identify an optimal annealing time, the
  last five linear sizes $L$ are consistent for a wide range of
  annealing times.}
\end{figure}

Figure (\ref{fig:tts_comp_scaling}) summarizes our results where
\DW is compared to both \HFS and \ICM. Interestingly, excluding the
region of small $\lambda$ where \ICML is designed to be the fastest,
\DW always performs better than the two classical heuristics for the
considered values of $\lambda$, being approximately $10^2$ times faster for
$\lambda = 7$. To better appreciate the different computational
scaling among the classical and quantum heuristics we analyzed,
Fig.~(\ref{fig:tts_comp_alpha_beta_icm_hfs}), top panels, reports
the scaling exponent $\beta$ of an exponential fit of the form:
\begin{equation}
  \log_{10} \text{TTS}(t_\text{run}) = \beta\,t_\text{run} + \alpha.
\end{equation}
In the plots, boxes represent the confidence interval $35\%$--$65\%$ for
$\beta$ computed using only the $50\%$ percentile of TTS while whisker
bars represent the confidence interval $35\%$--$65\%$ for $\beta$
computed using the $35\%$--$65\%$ percentile of the TTS. As one can see,
while \HFS is statistically indistinguishable from the \DW~data, \ICM performs
slightly better for large $\lambda$. However, the better performance for
\ICM for large $\lambda$ can be explained by noticing that \ICM has been
optimized for each $\lambda$ while both \DW and \HFS use the same setup
regardless of the value of $\lambda$.
Figure (\ref{fig:tts_comp_alpha_beta_icm_hfs}, bottom panels, shows
that \DW is consistently faster than both \HFS and \ICM by, on average,
a factor of $10^2$. Unfortunately, we cannot ``certify'' the \DW
computational scaling because we are not able to find the optimal
annealing time for the allowed minimum annealing time in the
device. Still, as shown in Fig.~(\ref{fig:tts_dw_varying_ann}),
we have strong indication that the computational scaling we have found
is reliable because of its stability for a large variation of annealing times
\cite{onemic}.

\begin{figure}
  \includegraphics[width=\columnwidth]{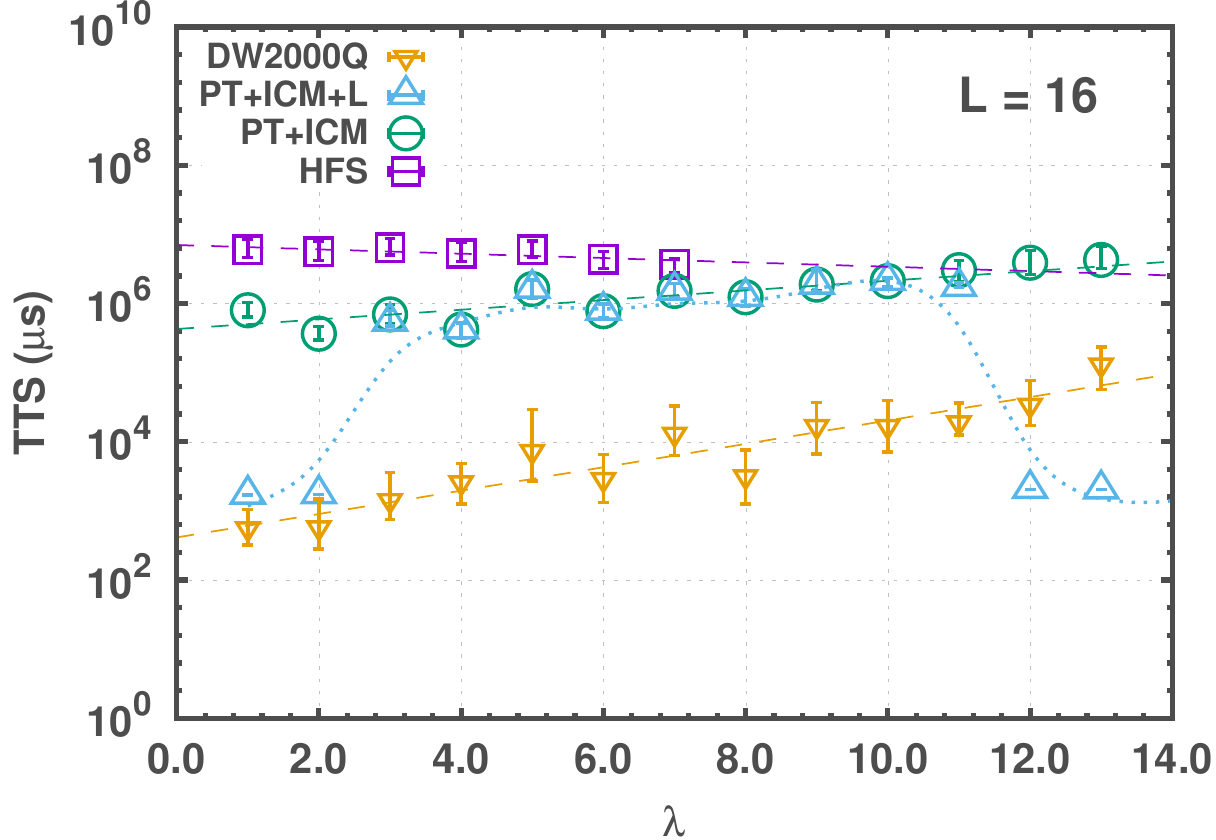}
  \caption{\label{fig:tts_comp_L16} Time-to-solution (TTS) for the
  parallel tempering isoenergetic cluster method (\ICMnsp), the Hamze-de
  Freitas-Selby (\HFSnsp) heuristic, as well for D-Wave 2000Q (\DWnsp)
  quantum chip.  All data points are for fixed linear lattice size $L =
  16$.  For this analysis, we also used a modified \ICM algorithm
  (\ICMLnsp) to take advantage of the knowledge of the virtual grounds
  state for both small ($\lambda \to 1$) and large ($\lambda \gg 1$)
  scaling limits (see main text for more details). Data show that \HFS
  is barely affected by the scaling $\lambda$, while the \DW performance
  slowly degrades by increasing $\lambda$ (most likely due to precision
  issues \cite{venturelli:15a, katzgraber:15}). As expected, \ICML
  performs the best for both small and large $\lambda$, while its
  performance quickly degenerates for $\lambda \sim 7$, i.e., in the
  region where the true ground state is different from the virtual
  ground state.  Interestingly, for $\lambda \sim 7$, \DW has the better
  performance resulting and is at least $100$ times faster than \HFS and
  \ICMnsp.}
\end{figure}

To further analyze the effects of the rescaling factor $\lambda$ in
Fig.~(\ref{fig:tts_comp_L16}) we show the performance of \DW compared to
\ICMnsp, \ICMLnsp, and \HFS at fixed linear size of the system $L$. As
expected, \ICML performs
the best for both small and large $\lambda$, while its performance
quickly degenerates for $\lambda \sim 7$, i.e., in the region where the
true ground state is a nontrivial overlap of the virtual ground states
at either small or large $\lambda$ [see
Fig.~(\ref{fig:fract_broken}) for the number of ``broken''
virtual variables for either the virtual planar model or the virtual
fully-connected bipartite model]. In contrast, the \DW performance
gradually decreases by increasing the scaling factor $\lambda$
(precision issues \cite{venturelli:15a, katzgraber:15} may be one of
the dominant factor of the loss of performance).

\begin{figure*}
 \includegraphics[width=0.9\columnwidth]{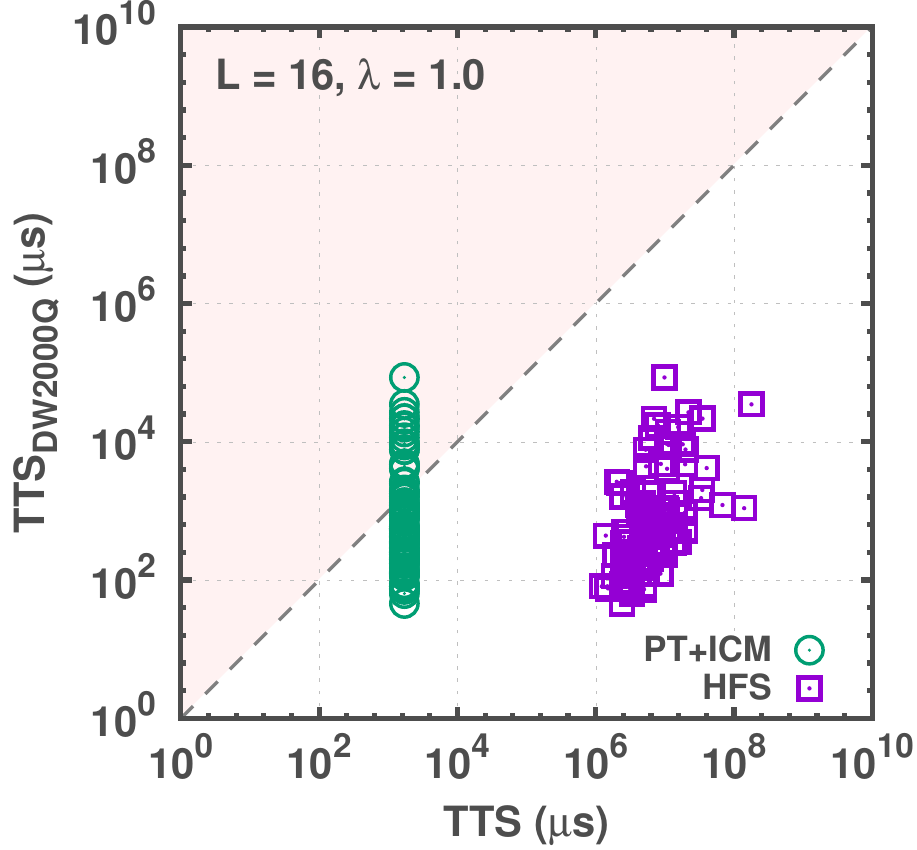}\hspace{1em}
 \includegraphics[width=0.9\columnwidth]{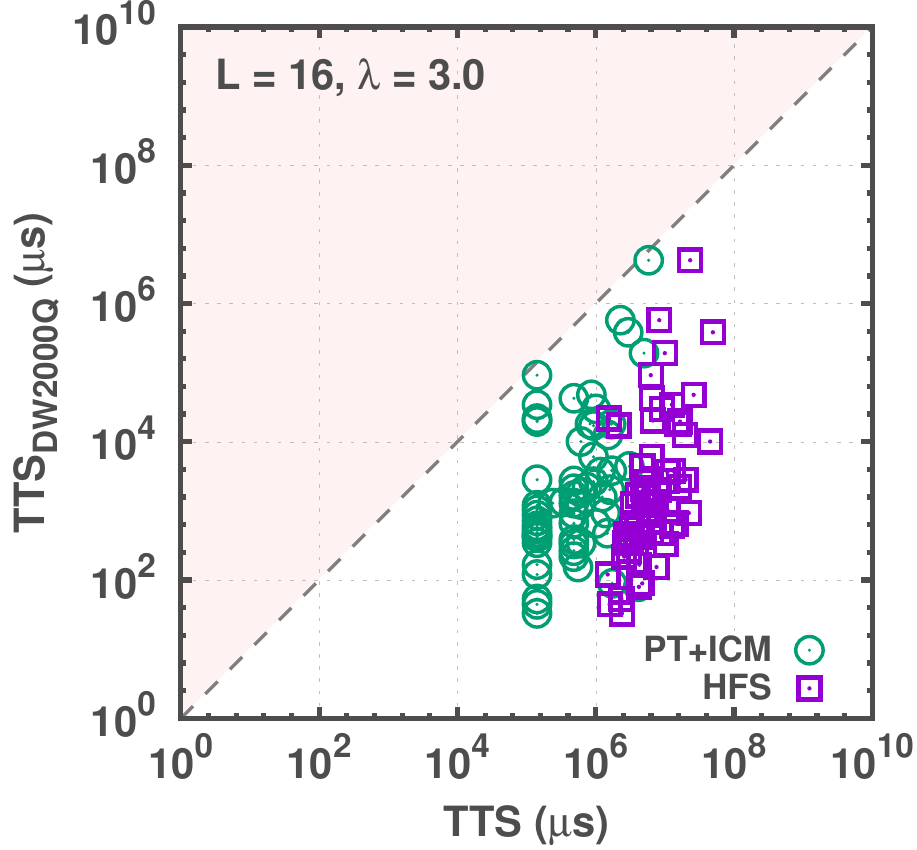}\vspace{1em}
 \includegraphics[width=0.9\columnwidth]{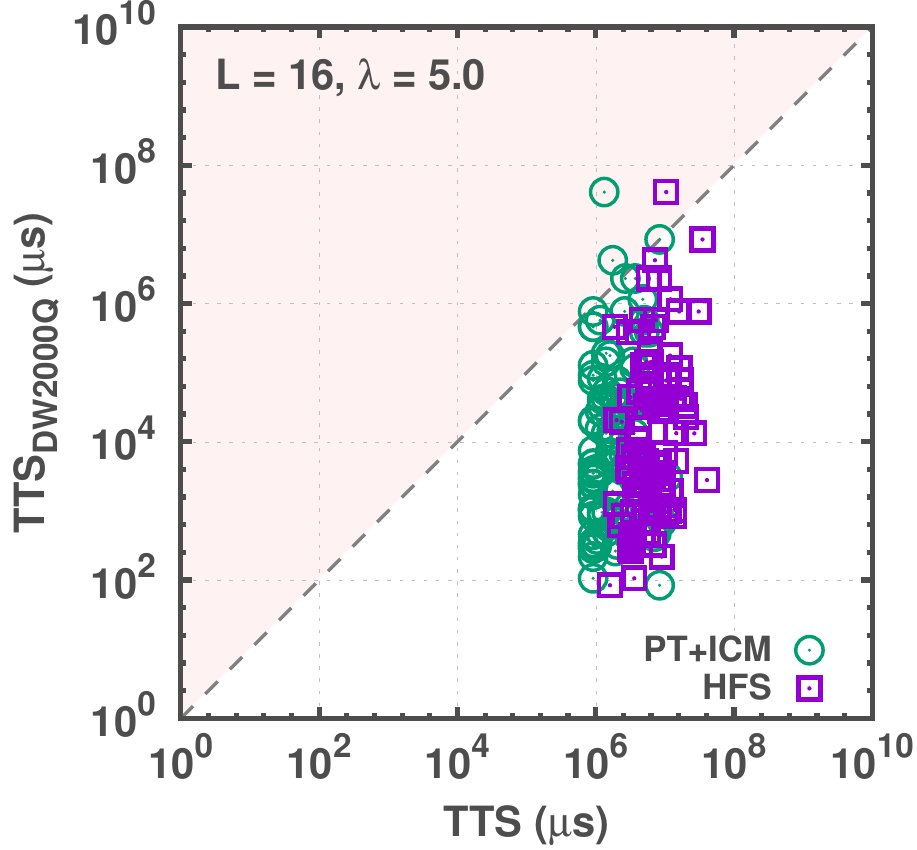}\hspace{1em}
  \includegraphics[width=0.9\columnwidth]{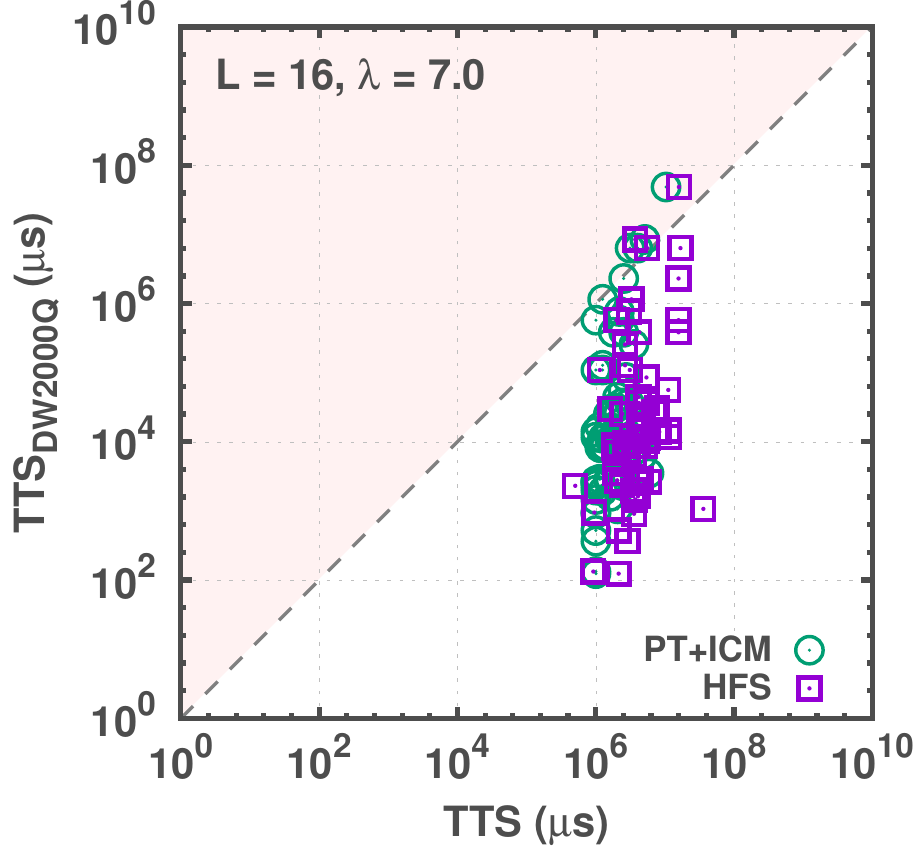}
  \caption{\label{fig:tts_comp_scatter} Instance by instance comparison
  of the Time-to-solution (TTS) between the different heuristics used,
  the parallel tempering isoenergetic cluster method (\ICM) and the
  Hamze-de Freitas-Selby (\HFS) algorithms, and the D-Wave 2000Q (\DWnsp)
  quantum chip. As one can see, \DW is consistently faster than \HFS and
  \ICMnsp.}
\end{figure*}

As final remark, it is important to stress that the advantage is not
just on the typical instance. Indeed, as it is shown in
Fig.~(\ref{fig:tts_comp_scatter}), \DW performs best even in an
instance by instance comparison, when contrasted to \HFS and \ICMnsp.

\section{Analysis of the Initialization and Readout Time}\label{sec:readout}

While the manufacturing of quantum hardware is becoming more advanced
with better superconducting circuits and more resilient qubits, quantum
technology, in general, is still in its infancy. Indeed, technological
problems such as long initialization and readout times are still a
burden for any early quantum device.  Hopefully, this problem will be
resolved in the near future with, e.g., cryogenic control electronics.
In particular, the \DW quantum chip needs $t_\text{read} = 123\,\mu s$
to read a configuration at the end of the annealing process. Moreover,
the \DW quantum chip requires $t_\text{ini} = 21\,\mu s$ between anneals
to reduce memory effects. In many cases, the total amount of
$t_\text{ini} + t_\text{read} = 144\,\mu s$ is an order of magnitude
larger than the optimal annealing time. Consequently, the inclusion of
the initialization/readout time affects more optimization runs that
involve fast anneals with many repetitions rather than slow anneals with
few repetitions.

\begin{figure*}
 \includegraphics[width=0.9\columnwidth]{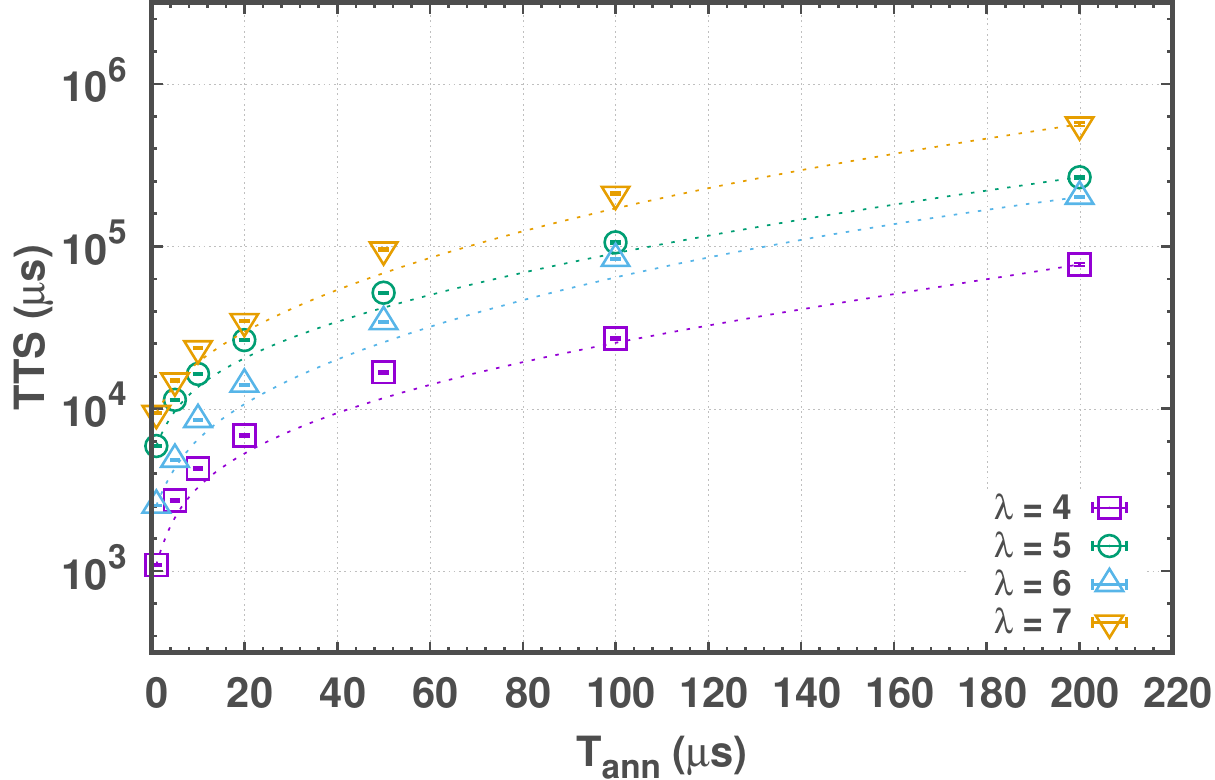}\hspace{1em}
 \includegraphics[width=0.9\columnwidth]{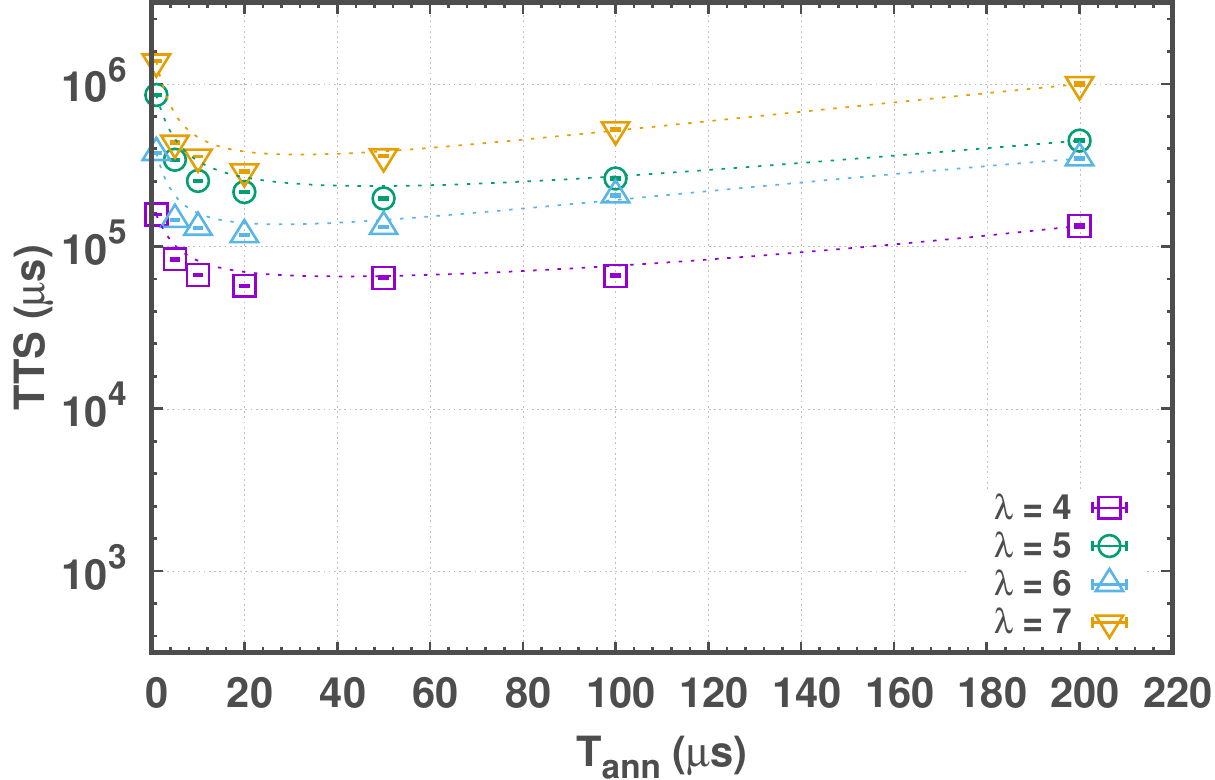}
 \caption{\label{fig:tts_comp_with_ini}Time to solution (TTS) by varying the
 annealing time $T_\text{ann}$ without (left panel) and with (right panel) the
 inclusion of the initialization and readout time. Interestingly, the inclusion
 of the initialization and readout time moves the optimal
 annealing time to larger values.}
\end{figure*}

\begin{figure}
  \includegraphics[width=0.9\columnwidth]{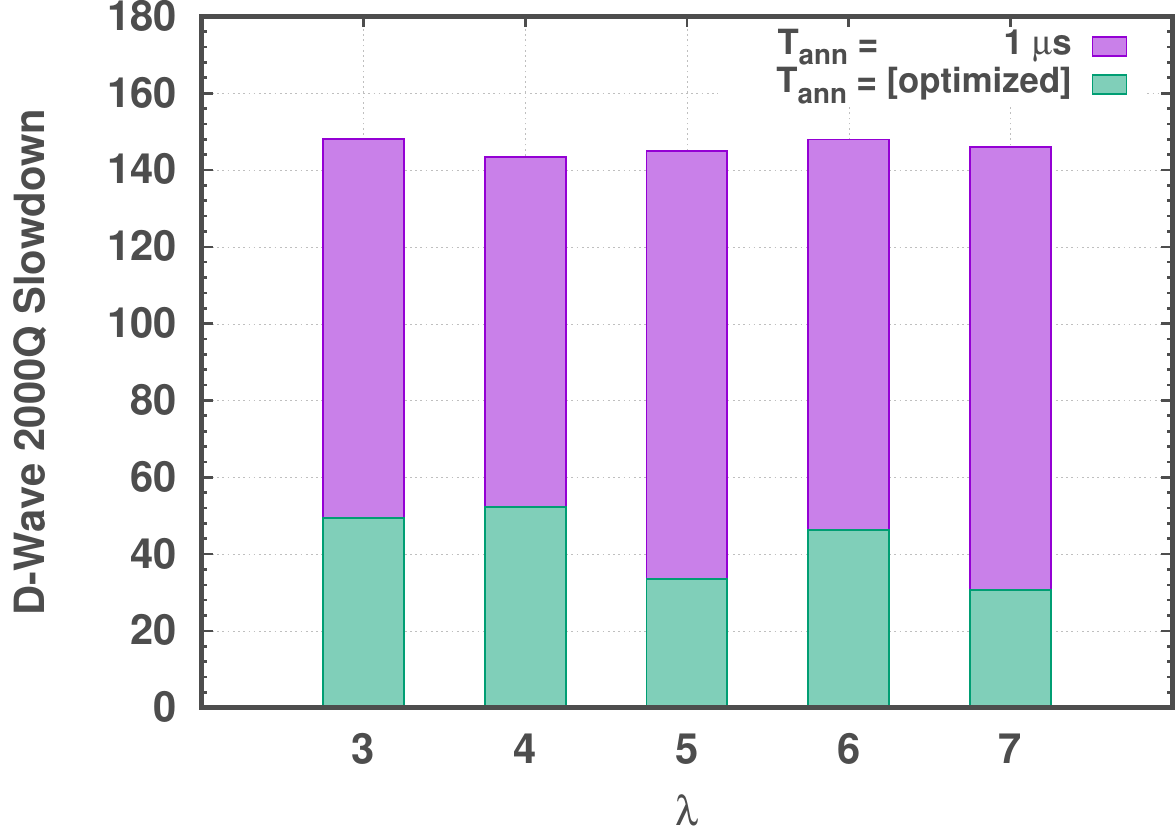}\hspace{1em}
  \caption{\label{fig:slowdown_with_ini}Ratio between the time to solution (TTS)
  before and after the inclusion of the initialization and readout time
  (slowdown) for $L = 16$. As one can see, the optimization of TTS with the
  inclusion of the initialization and readout time drastically reduces the
  slowdown of the \DWnsp.}
\end{figure}

As described in Sec.~\ref{sec:results}, the optimal annealing time for
the DCL class of problems is $1\,\mu s$ (see
Fig.~\ref{fig:tts_comp_with_ini}, left panel).  Therefore, the addition
of $144\,\mu s$ drastically reduces the performance of the \DW quantum
chip hence limiting the quantum advantage shown in
Fig.~\ref{fig:tts_comp_L16}. The optimization described in
Sec.~\ref{sec:results} does not take into account the initialization and
readout times. A more careful analysis shows that the inclusion of the
initialization and readout times shifts the optimal annealing time to
larger values (see Fig.~\ref{fig:tts_comp_with_ini}, right panel).
Indeed, once the initialization and readout times are included in the
optimization, the overall performance is improved by having slower
anneals with fewer repetitions.  Interestingly, as shown in
Fig.~\ref{fig:slowdown_with_ini}, the overall slowdown due to the
introduction of the initialization and readout times can be reduce by a
factor $\sim\!140$ to a mere factor $\sim\!30$, which allows the \DW
device to keep a small computational advantage in comparison to the \HFS
and \ICM heuristics.

\section{Inclusion of Power Consumption}\label{sec:discussion}

While most of the benchmark studies have largely focused on pure
computational speed, the inclusion of power consumption has been largely
neglected in the literature \cite{power}. With ever-growing data
centers, power consumption has become an important issue and ``greener''
computational solutions are highly sought after.

A large-sized data center like the one hosted at NASA Ames
\cite{top500,nasa} has a typical energy consumption of approximately
$5$MW, with a $4:1$ ratio between power usage and cooling (during
benchmarking).  With more than $245\,000$ cores for the current NASA
high-performance computing cluster, the typical energy consumption is
approximately $20$W/core.  In contrast, the energy consumption of the
\DW quantum processing unit is approximately $150$pW. Keeping the
quantum processing unit cooled to $20$mK requires approximately $15$
-- $25$kW.  In our analysis, we found that the \DW device was
approximately $10^2-10^3$ times faster than the used \ICM and \HFS
heuristics.  Therefore, to compete against \DWnsp, $\sim 10^2$ --
$10^3$ compute cores are needed running in parallel with a total
energy consumption between $2$ and $20$kW. Therefore, power
consumption is, overall, comparable. However, there is a remarkable
difference: The data center uses $80\%$ of the total consumed energy
to run the computers, while the \DW device requires only $10^{-14}\%$
of the power to run the quantum processing unit.  Therefore, while an
improvement of the power usage effectiveness (PUE) \cite{brady:13,
greengrid} for the classical data center would eventually reduce the
total cooling power of $20\%$, far more efficient cooling alternatives
are needed to reduce the quantum PUE (qPUE). It is unclear how the
qPUE can be reduced due to the cryogenic requirements for quantum
processing units. However, dry dilution refrigerators with more
efficient pumping systems might improve this metric \cite{power2}.

\section{Conclusions}\label{sec:conclusions}

In conclusion, we present the first class of tunable benchmark problems
-- Deceptive Cluster Loops (DCL) -- for which the D-Wave quantum chip
(\DWnsp) shows an advantage over the currently best classical
heuristics, namely the parallel tempering isoenergetic cluster method
(\ICMnsp) and the Hamze-de Freitas-Selby (\HFSnsp) algorithm. The
benchmarks are characterized by a control parameter $\lambda$, the
scaling factor of the inter-cell couplings, that allows to continuously
transform the model from a virtual planar model ($\lambda \sim 1$) to a
virtual fully-connected bipartite problem ($\lambda \gg 1$). While
classical heuristics are faster in the small- and large-$\lambda$ limit
where the logical structure can be exploited, \DW is the fastest in the
crossover region $\lambda \sim 7$, where the DCL problems are neither
virtual planar nor virtual fully-connected bipartite. Indeed, while the
computational scaling is comparable among classical and quantum
heuristics, the \DW~device is approximately two orders of magnitude
faster than the currently best known heuristics (\ICM and \HFSnsp) with
a comparable scaling. This result represents the first of its kind since
the inception of the D-Wave quantum chip.  We also show that the
inclusion of the initialization and readout times may reduce the
performance of the \DW quantum chip. However, we also show that a proper
optimization of the time to solution mitigates the effects of the
inclusion of the initialization and readout times, allowing the \DW
quantum chip to have a small advantage over the best classical
heuristics to date.

\section{Acknowledgments}\label{sec:ackno}

H.~G.~K.~acknowledges support from the National Science Foundation
(Grant No.~DMR-1151387) and thanks M.~Thom for multiple discussions on
power consumption of the \DW devices. He also thanks N.~Artner for
support. S.~M.~acknowledges E.~G.~Rieffel for the careful reading of the
manuscript and useful discussion, C.~E.~Henze and the NASA Ames Research Center for
support and computational resources.  We would like to thank
M.~Steininger and J.~King for feedback on the manuscript.  This research
is based upon work supported by the Office of the Director of National
Intelligence (ODNI), Intelligence Advanced Research Projects Activity
(IARPA), via Interagency Umbrella Agreement IA1-1198. The views and
conclusions contained herein are those of the authors and should not be
interpreted as necessarily representing the official policies or
endorsements, either expressed or implied, of the ODNI, IARPA, or the
U.S.~Government.  The U.S.~Government is authorized to reproduce and
distribute reprints for Governmental purposes notwithstanding any
copyright annotation thereon.

\appendix

\section{Number of Broken Virtual Variables in the DCL Model}

\begin{figure*}
  \includegraphics[width=\columnwidth]{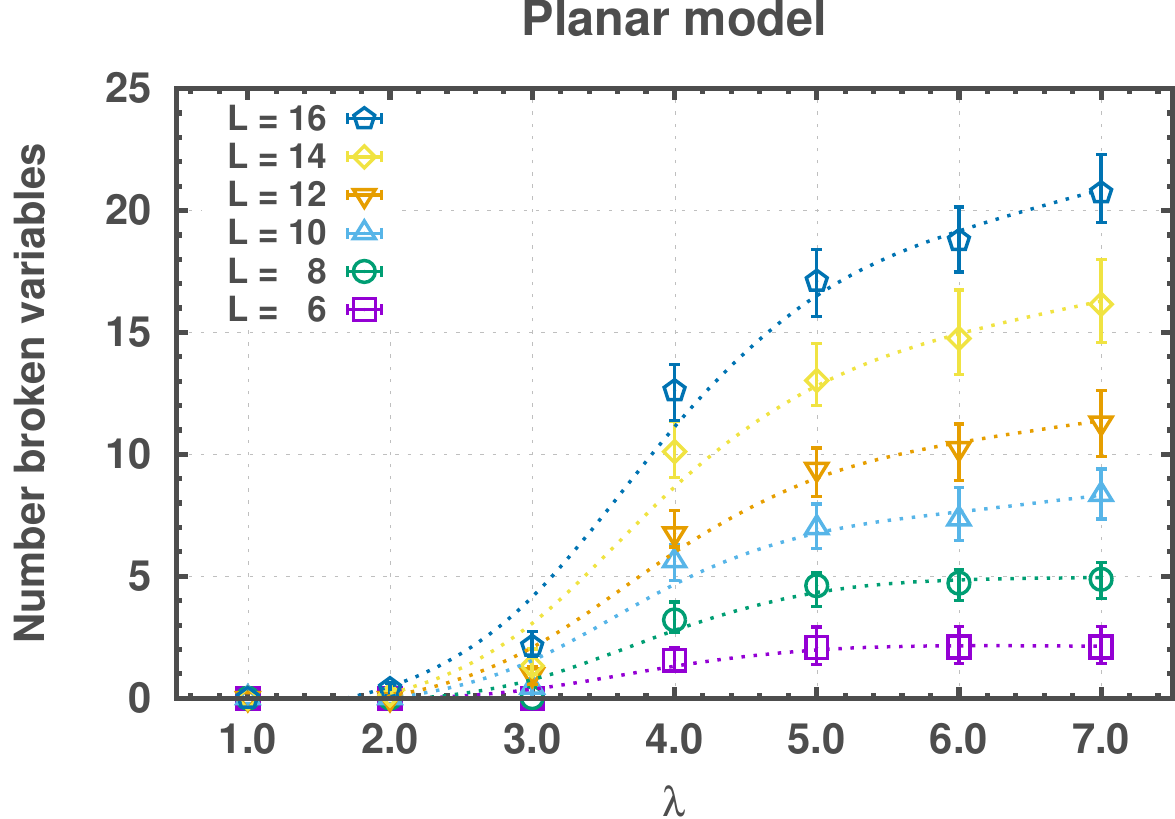}\hspace{1em}
  \includegraphics[width=\columnwidth]{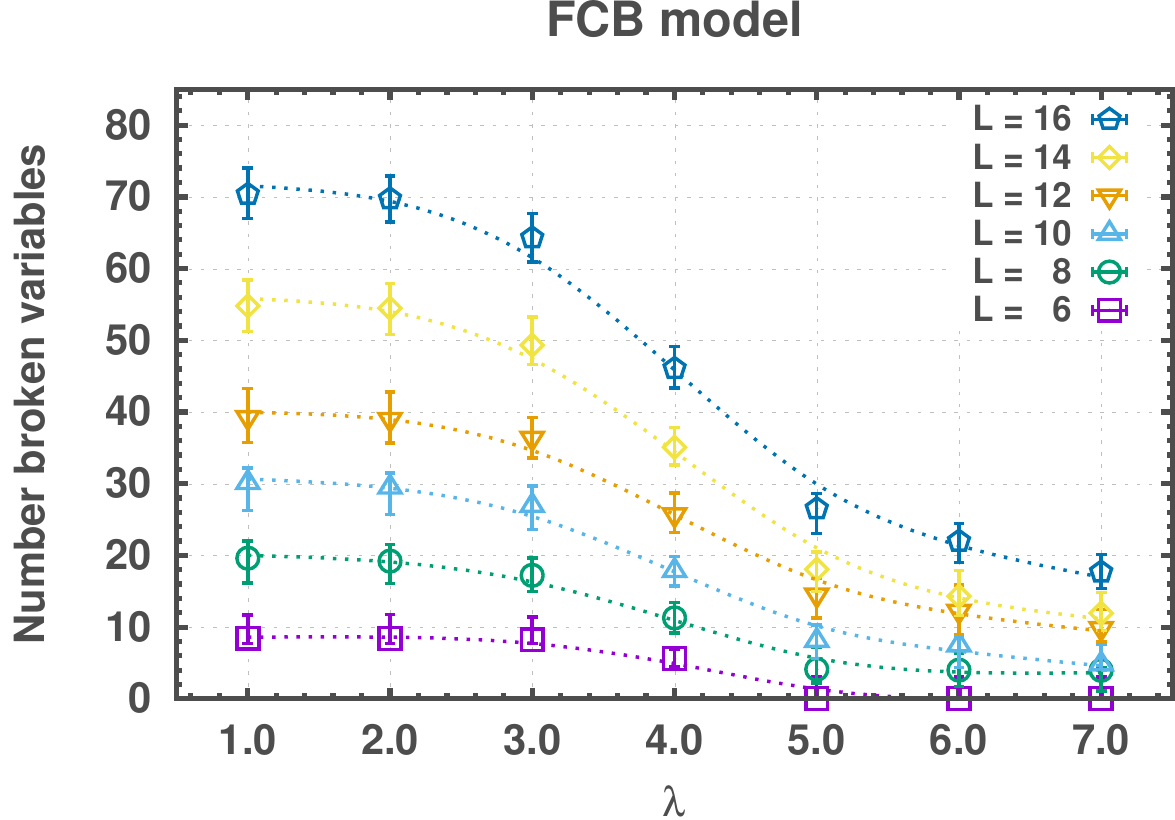}
  \caption{\label{fig:fract_broken} (Left panel) Number of broken
  virtual variables with respect to the virtual planar model
  (limit for $\lambda \to 1$). (Right panel) Number of broken
  virtual variables with respect to the virtual fully-connected
  bipartite model (limit for $\lambda \gg 1$). As one can see, the
  number of broken variables goes to zero when the scaling $\lambda$
  approaches the limit of the corresponding virtual model.}
\end{figure*}

Our numerical simulations for $L = 16$ suggest that the DCL model
reduces to the virtual planar model for $\lambda \lesssim 2$, while it
is a virtual fully-connected bipartite for $\lambda \gtrsim 10$ [see
Fig.~(\ref{fig:tts_comp_L16})].
Figure (\ref{fig:fract_broken}) shows the number of ``broken''
virtual variables for either the virtual planar model or the virtual
fully-connected bipartite model.  The interesting regime is obtained
when $\lambda \sim 7$, i.e., when the DCL model is neither a virtual
planar model nor a virtual fully-connected bipartite model. In this
regime, the ground state of the DCL model cannot be found by solving a
corresponding virtual problems and therefore, logical structures
cannot be exploited as in Refs.~\cite{king:15,king:17,mandra:17a}.

\section{Simulation parameters}

In this Section, we briefly report the main parameters we used
for our experiments and numerical simulations.\\

\emph{DCL Random Instances} --- We randomly generate $100$ instances for
each system size ($L = 6, \ldots,\, 16$). The instances are generated by
following the prescription in Ref.~\cite{king:17} (with $\alpha = 0.24$,
$R = \rho = 1$) and then properly rescaling the inter-cell couplings by
a factor $\lambda$.
To make the problem easier and allow for a better analysis of the computational
scaling, we break the loops removing one of the ferromagnetic couplings.
For consistency, the same instances have been used for all values of
$\lambda$. Unlike in Ref.~\cite{king:17}, we used all available
qubits, i.e., some of the unit cells are not complete.\\

\emph{\DW Parameters} --- For all experiments, we use the minimum
allowed annealing time of $5\,\mu s$ and $100$ gauges $\times$ $1000$
runs, i.e., $100\,000$ total readouts. The initialization time and the
readout time have not been included in the calculation of the TTS.\\

\emph{\ICM Parameters} --- The lowest and highest temperature for
parallel tempering have been chosen to be $1$ and $10$, respectively, to
maximize the performance of \ICMnsp. Optimal sweeps for each instance
and $\lambda$ have been determined by computing the cumulative
distribution of the probability to find the ground state ($100$ runs
for each instance and $\lambda$). The overall optimal number of sweeps is
then obtained by bootstrapping the optimal number of sweeps for each
instance. For all simulations, the number of sweeps has been
optimized to minimize the TTS of the $50\%$ percentile. The
initialization time and the readout time have been not included in the
calculation of the TTS.\\

\emph{\ICML Parameters} --- The parameters used are the same as for
\ICMnsp. The computational time to find the ground state of either the
virtual planar model of the virtual fully-connected bipartite model has
been set to zero (in reality, the computational time to find the ground
state of the fully-connected bipartite model is nonnegligible).\\

\emph{\HFS Parameters} --- The option -S13, namely ``Exhaust maximal
tree width 1 subgraphs'' with partial random state initialization, has
been used.  Optimal sweeps for each instance and $\lambda$ have been
determined by computing the cumulative distribution of the probability
to find the ground state ($100$ runs for each instance and $\lambda$).
The overall optimal number of sweeps is then obtained by bootstrapping
the optimal number of sweeps for each instance. For all the simulations,
the number of sweeps has been optimized to minimize the TTS of the
$50\%$ percentile. The initialization time and the readout time have
been not included in the calculation of the TTS.\\

\begin{figure*}
  \includegraphics[width=0.48\textwidth]{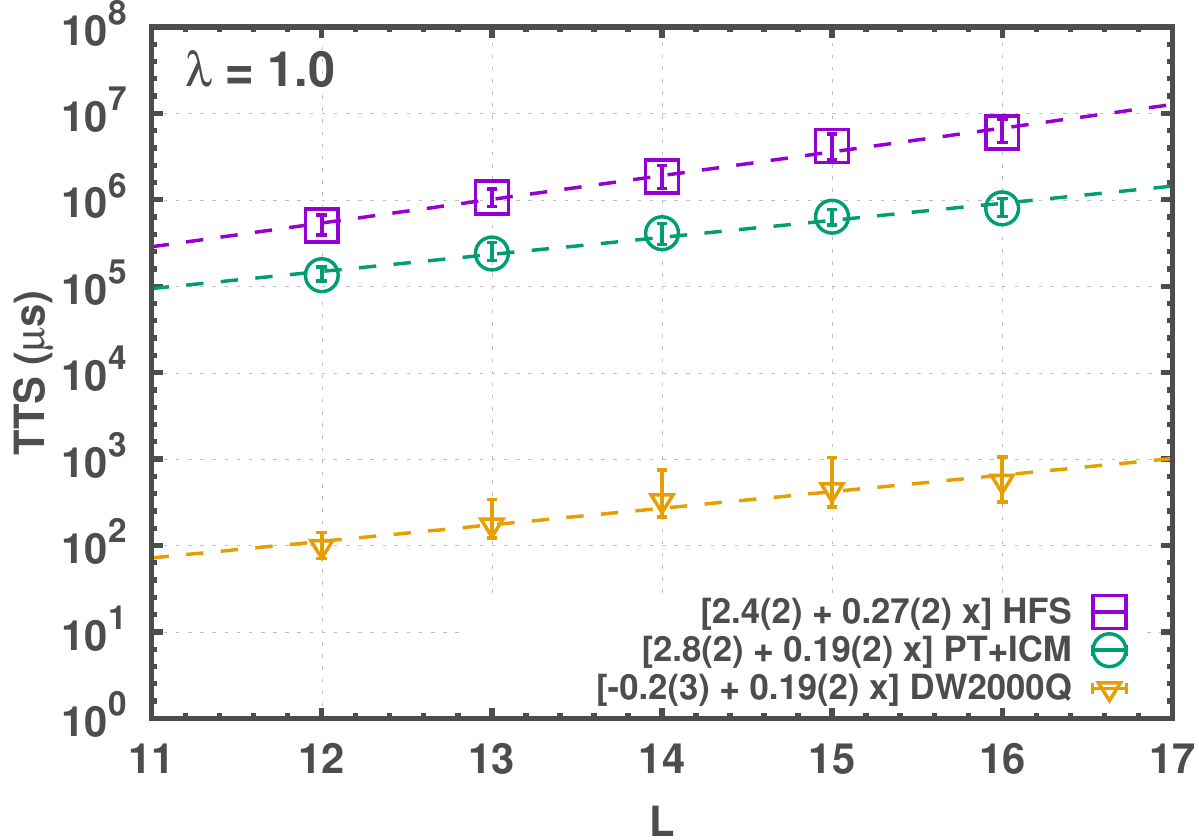}\hspace{1em}
  \includegraphics[width=0.48\textwidth]{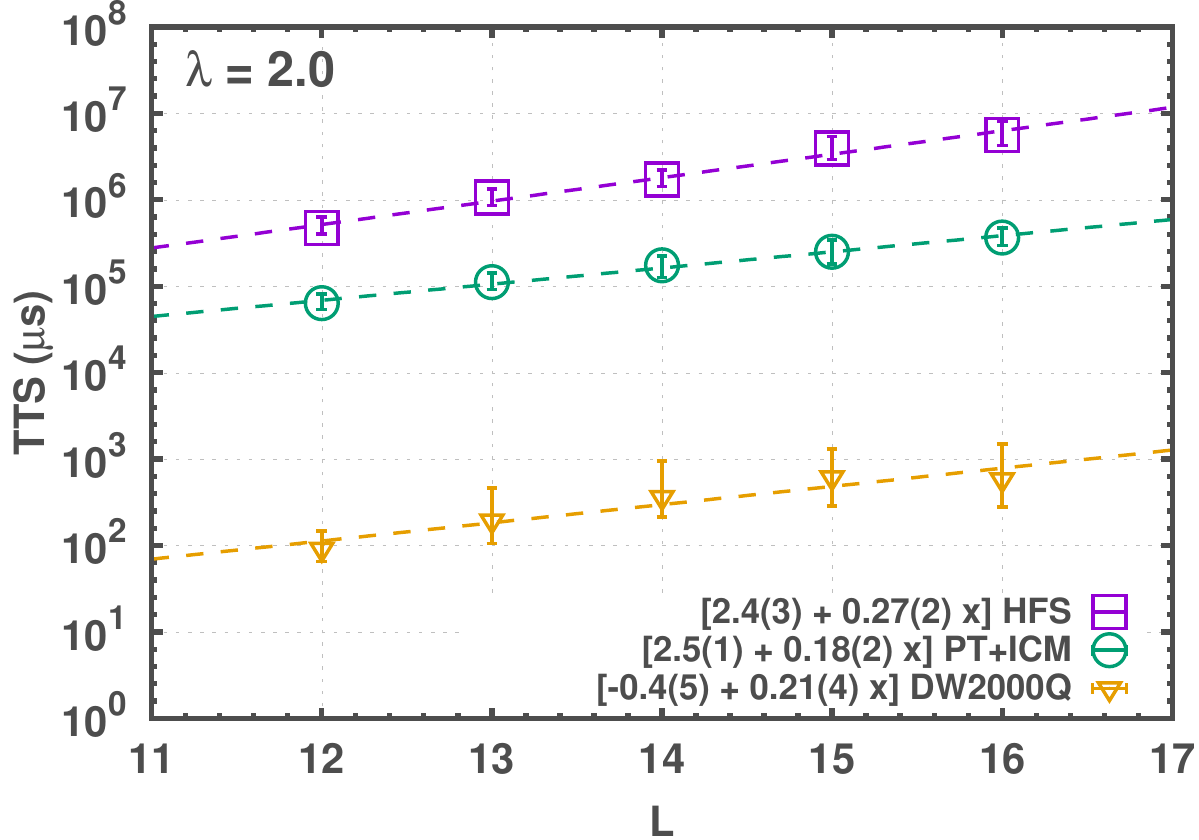}\vspace{1em}\\
  \includegraphics[width=0.48\textwidth]{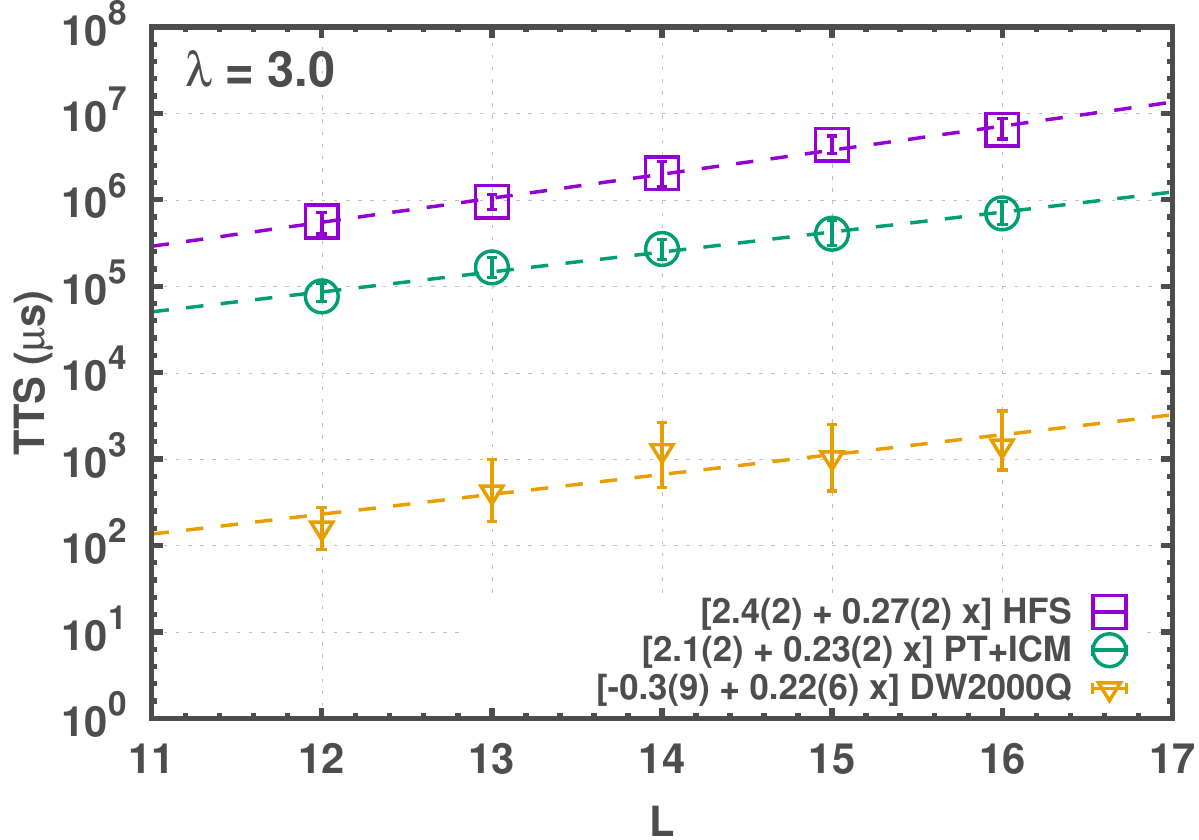}\hspace{1em}
  \includegraphics[width=0.48\textwidth]{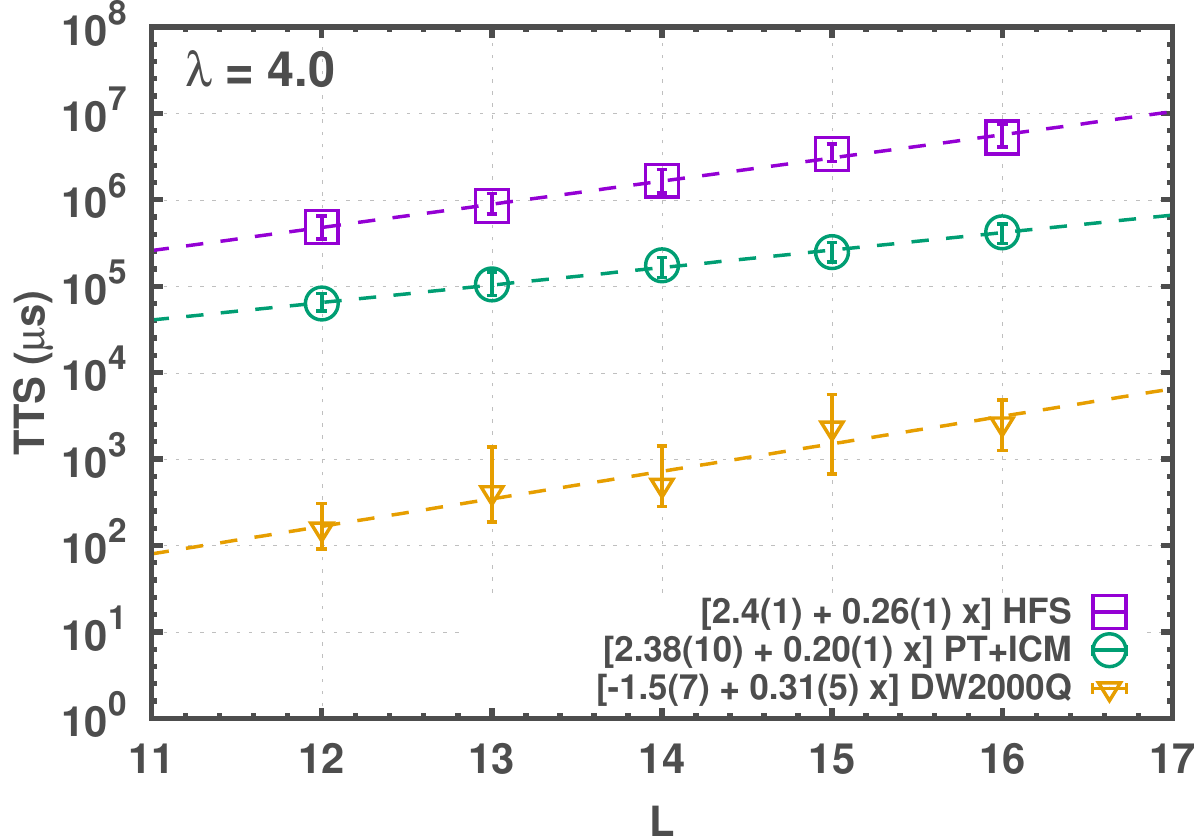}\vspace{1em}\\
  \includegraphics[width=0.48\textwidth]{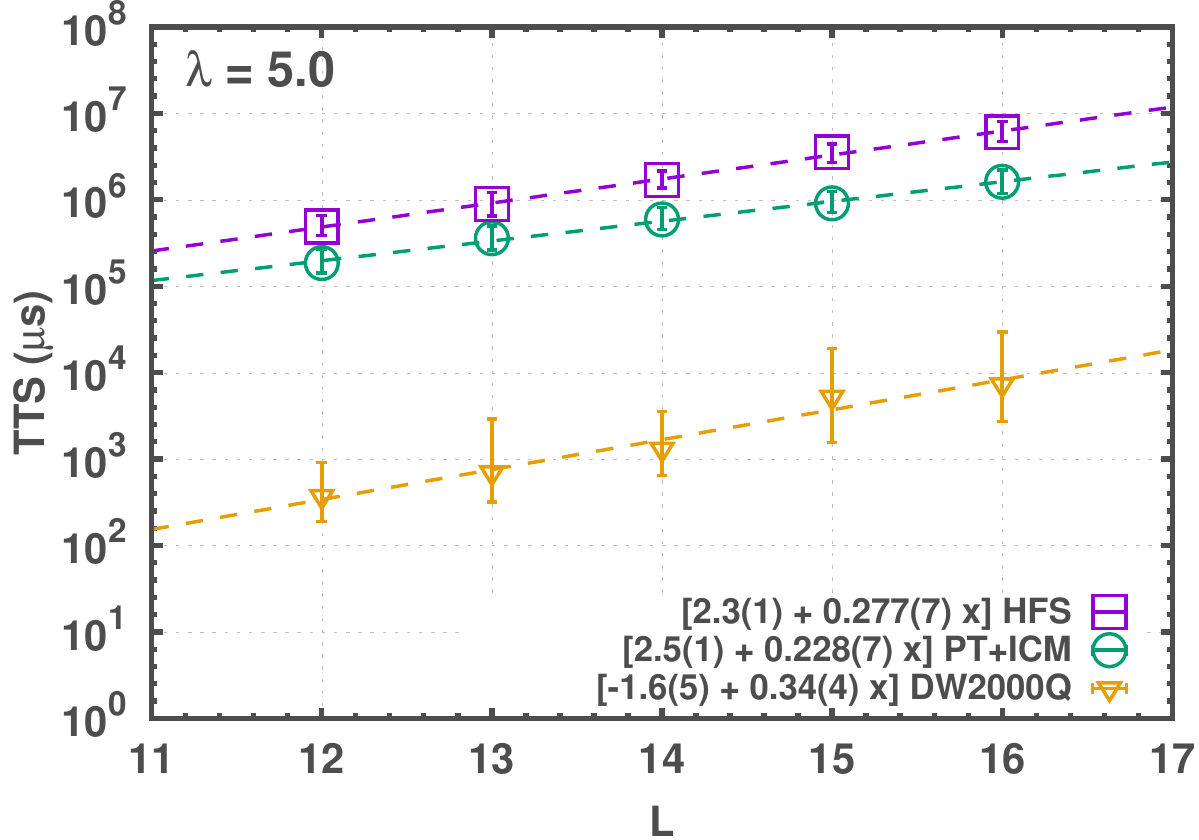}\hspace{1em}
  \includegraphics[width=0.48\textwidth]{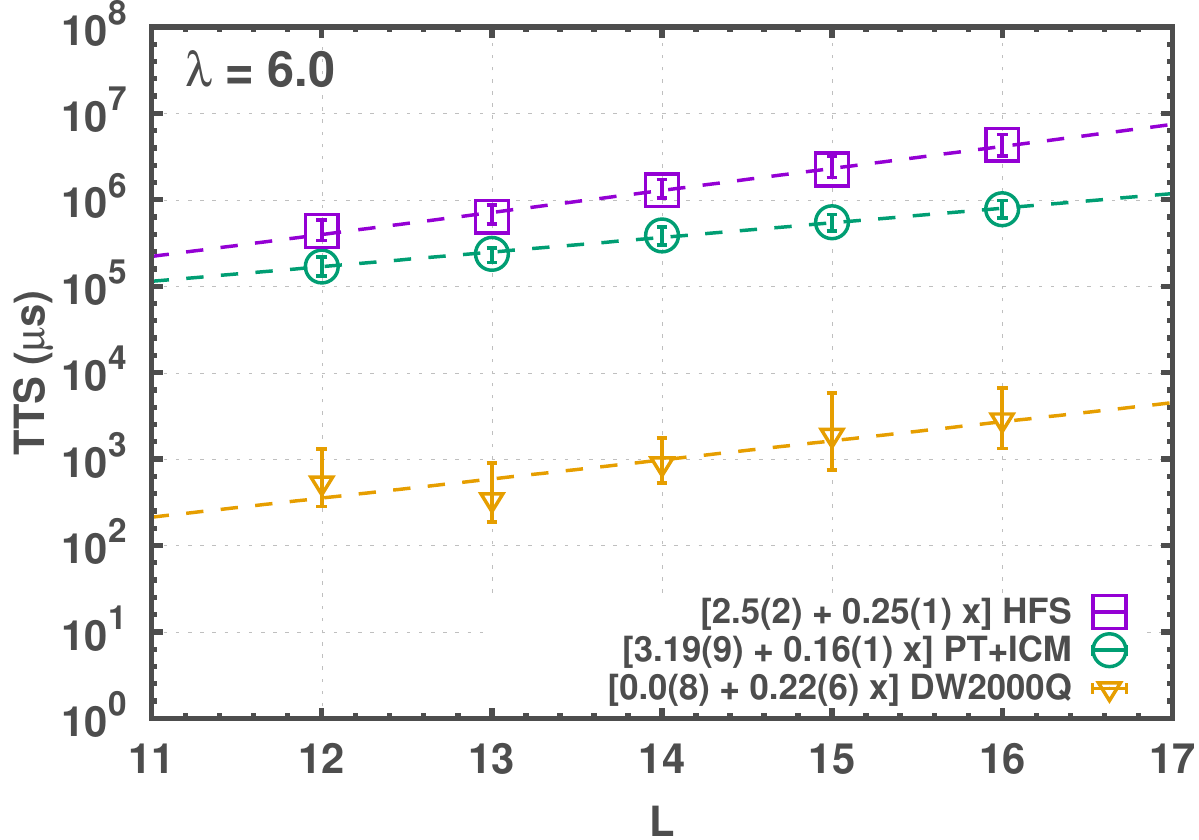}\hspace{1em}
  \caption{\label{fig:tts_slope_scaling_all}Linear regressions by
  varying the linear system size $L$, at different inter-cell
  couplings scaling $\lambda$.}
\end{figure*}

\emph{Exponential Fits} --- For all the linear regressions, only the
last $5$ system sizes ($L = 12,\, \ldots,\, 16$) have been used.
$\log_{10} \text{TTS}$ fits for the $50\%$ percentile have been obtained
using a linear least squares model. $\log_{10} \text{TTS}$ fits for the
$35\%-65\%$ confidence have been computed by randomly extract values in
the confidence interval, one for each size, and then bootstrapping the
linear regression data.  Figure (\ref{fig:tts_slope_scaling_all}) shows
the fits for different values of $\lambda$.

\bibliography{refs,comments}

\begin{thebibliography}{47}
\bibitem[{chi()}]{chicken}
\bibinfo{note}{{{It has been tempting to refer to the approach developed in
  Ref.~\cite{king:17} ``frustrated cluster loops Hen {\`a} la King,'' but that
  would be a mouthful every time it appears in the text.}}}
\bibitem[{sqa()}]{sqa}
\bibinfo{note}{{{An analysis using SQA has not been included in this study
  because preliminary results on the instances studied in Ref.~\cite{albash:17}
  show that PT+ICM is at least one order of magnitude faster than SQA with a
  comparable scaling.}}}
\bibitem[{pow({\natexlab{b}})}]{power2}
\bibinfo{note}{{{We note that a comprehensive analysis of the power consumption
  requires the inclusion of embedding, programming, and readout times, as well
  as considerations on dissipated energy, for both classical and quantum
  machines.}}}
\end{thebibliography}

\end{document}